# PSI Draft Specification

Mark Reid, James Montgomery, Barry Drake, Avraham Ruderman

Version 2. 12 September 2013

# 1 Introduction

The *Protocols and Structures for Inference* (PSI) project aims to develop an architecture for presenting machine learning algorithms, their inputs (data) and outputs (predictors) as resource-oriented RESTful web services [1] in order to make machine learning technology accessible to a broader range of people than just machine learning researchers.

Currently, many machine learning implementations (e.g., in toolkits such as Weka, Orange, Elefant, Shogun, SciKit.Learn, etc.) are tied to specific choices of programming language, and data sets to particular formats (e.g., CSV, svmlight, ARFF). This limits their *accessability*, since new users may have to learn a new programming language to run a learner or write a parser for a new data format, and their *interoperability*, requiring data format converters and multiple language platforms. To address these limitations, the aim of the PSI service architecture is to present the main inferential entities – relations, attributes, learners, and predictors – as *web resources* that are accessible via a common interface. While there is a growing number of existing machine learning web services, each has its own API and is tailored to suit a different subset of machine learning activities. By providing a consistent interface for the entities involved in learning, the PSI approach can improve interoperability and hide implementation details to promote accessibility. This will allow PSI to support *federated* machine learning solutions composed from multiple, independent services.

The purpose of this document is to specify how learning algorithms, data, and predictors can be presented as RESTful web resources. Learning algorithms, data, predictors, and other resources conforming to this specification will be collectively referred to as *PSI services*. Although examples are used in this specification, this document is *not* intended as an introduction to or tutorial for programming against PSI services. Tutorials will be made available separately. Also, this document explicitly does *not* specify how to implement any particular learning algorithms or data sets as PSI services. Reference libraries and example code for implementing PSI services may be released separately as part of the PSI project.

There are two intended audiences for this specification: *providers* of PSI services and *consumers* of PSI services. Providers may include machine learning researchers who wish to present their data and learning algorithms as PSI services for others to use. Consumers may include developers who have little background in machine learning but wish to build predictive models for data they may have. We assume both audiences have some background in the use of RESTful web service APIs. In particular, we assume some familiarity with the HyperText Transfer Protocol (HTTP) [2] and the JavaScript Object Notation (JSON) [3,4].

This document is organised as follows. The remainder of this introduction describes what machine learning problems are, the design decisions that were made to model these problems, and brief overview of values and schema. Section 2 details the resulting architecture in terms of the main PSI resource types: transformers, relations, attributes, learners, and predictors. Two key features of the architecture are explained in detail: the use and composition of *attributes* – functions for extracting values from data – to enable configurable structured data representation; and a simple but flexible *schema language* for describing and prescribing the structures used to represent data. The purpose of these two features is to make it easy for human or machine consumers of PSI services to "glue" services together, that is, to readily determine what input representation a PSI service requires and construct a conforming representation from the output of other PSI services. Section 3 presents an example of how a hypothetical implementation of PSI specification might be used to solve a simple machine learning problem. Section 4 outlines planned revisions to the specification. Appendix A gives details of the PSI schema language and Appendix B defines a number of useful schema that are provided with any PSI implementation.

## 1.1 Modelling Machine Learning Problems

There are a diverse range of problems within machine learning that can be cast in a *data-learner-predictor* framework. These include classification, regression, density estimation, ranking, dimensional reduction, collaborative filtering, structured prediction, and more. As the name of their grouping suggests, each of these problems can be solved by the application of a learning algorithm to a data set of instances to produce a predictor that can make inferences about additional, previously unseen instances.

The following three scenarios are representative of three common types of machine learning problems – classification, ranking, and probability estimation – and differ in several important ways.

**Iris Classification**

R.A. Fisher's famous problem of categorising irises [5] into one of three species (*setosa*, *virginica* and *versicolor*) based on measurements of their petal and sepal dimensions is an example of a simple classification problem. The canonical data set for this problem consists of 150 instances represented by feature vectors with four real-valued measurements and a species label.

An alternative way of presenting this problem would be to classify *photographs* of the irises in question, thus saving the time needed to manually measure features of the examples (especially useful when applying machine learning in the field). Each instance would then have a single feature, its image data encoded in JPEG format, which is used in both training (together with a species label) and later classification. Internally, a learning algorithm may derive additional features describing shape and colour, but only the original JPEG image is required as input.

**Real Estate Ranking**

Properties in a housing market are described using three features: their suburb name, the number of bedrooms, and their land area. Examples of a buyer's preferences for properties are available as pairs of houses described by their features, the first being preferred to the second. A learner for this problem is required to infer a scoring function for houses (including new, unseen ones) based on their features after being shown example preferences. The scores are used to construct a total order over houses, with higher scores indicating more preferred houses.

**Document Class Probability Estimation**

Documents from a corpus are described using a "bag of words" representation: a map from words appearing in the document to their term frequency–inverse document frequency (TF-IDF) weight. Given a collection of documents, each labelled as "spam" (i.e., junk), "ham" (ordinary), or "bacon" (important), a learner is required to estimate the probability over these three categories for new documents.

These three problems all exhibit different prediction types – categories, ranks, and probabilities – and different input structures – numeric feature vectors, pairs of mixed feature vectors, and bag of words. One aim of the PSI project is to provide a framework that is flexible enough to model all of these scenarios in a simple, consistent manner.

## 1.2 Design Decisions

This section outlines the reasoning behind the key design choices that were made during the development of this specification. At a high-level the main tension was between the desire to develop an architecture that is *flexible* enough to allow many different types of machine learning problems to be modelled (including the examples above), and *simple* enough so as not to dissuade people from implementing against it.

### 1.2.1 Choice of Architectural Style

Due to its conceptual elegance, sound principles, widespread adoption, and ease of interoperability with other web services, **the PSI service specification is designed using a Resource-Oriented Architectural style** (ROA). In this RESTful style, every interaction between a client and a server can be described as a service *Responding* to a *Request* with a *Representation* of a *Resource*. By adopting this style, several constraints must be adhered to, which influence the decisions below. These include the identification of resources using URIs (Uniform Resource Identifiers), the use of a small number of request modes (GET, POST, PUT, DELETE), the inspection and manipulation of resources through their representations, and the management of client-server state through linked resources.

Details about the Resource-Oriented Architectural style can be found in the book *RESTful Web Services* by Richardson and Ruby [1].

## 1.2.2 Choice of Resources

In the data-learner-predictor style of machine learning problems that this specification aims to formalise, solutions to learning problems are typically described in terms of the relations, learners, and predictors involved. These are therefore natural choices to present as services. Solving machine learning problems also frequently involves data *transformations*, of which prediction can be viewed as a subset. Thus the notion of a predictor is generalised to a *transformer*. Further, as described below, the communication between these services relies heavily on *attributes* and *schema* which are composed easily when presented as services. Thus, **relations, learners, transformers and predictors, attributes, and schema are resources within the PSI framework**.

Details of these resources can be found in Section 2 below.

## 1.2.3 Choice of Instance Representation

In ROA architectures, resources are never accessed directly. Instead, *representations* of the resources are used to form requests and responses to and from these resources. The "common currency" shared by PSI resources are instances: relations are collections of instances of the same "shape"; learners take relations as input to create predictors; predictors take instances as input to make predictions. In order for these resources to interact they must have a shared representation for instances. In the PSI framework this shared representation is JSON – the JavaScript Object Notation [3,4]. This is a widely used, well-supported, lightweight, easy-to-parse data structure that is rich enough to model many common machine learning data types such as dense and sparse feature vectors and matrices, probability distributions, sets, graphs, and text. For this reason, **all instances within the PSI service architecture are represented as JSON values**.

Details of the JSON specification and syntax can be found at `http://json.org`.

## 1.2.4 Use of Attributes and Schema

The variety of data structures used by machine learning algorithms and the flexibility of JSON for modeling them means there is no single choice of JSON structure that can be used for every learning problem. For example, a support vector machine algorithm may require instances to be represented as feature vectors of real numbers; an algorithm for classifying documents might require that documents are represented as "bags of words"; and a ranking algorithm may require pairs of mixed-type feature vectors. This variety leads to two important design constraints:

1. It must be possible to easily *construct* novel structures for representing instances for many different types of learner;
2. Learners must be able to *describe* the structure of the instance representation they require and attributes must be able to *describe* the structure of the values they emit.

The first constraint is met in the PSI framework through the use of **attributes**. These can be thought of as functions that map instances to values. Like JSON values, attributes can be composed through array and object structures to create new attributes that return more complex values. For example, an attribute that returns an integer and another that returns a string may be composed into a attribute that returns arrays containing one integer and one string. This flexibility is the reason that **all instance representations in the PSI service architecture are described using attributes**.

Details of attributes and their composition rules can be found in Section 2.6.

Because instance values are accessed via their attributes, and given the wide variety of alternative data formats and mechanisms for transferring data from clients to servers, data upload (i.e., creation of new relations) is deliberately excluded from this specification. Each PSI service is free to offer its own support for creating relations, and may even choose not to support user creation of data sets. Once a relation exists as a PSI resource, all further interaction with it takes place through its descriptive representation and its attributes, which hide the underlying data format used.

Describing the range of values an attribute can produce or a learner can take as input can be done through the use of a **schema language**. Each schema within a schema language describes a range of values through a set of constraints on the structure those values can have and the range of values within that structure. A value that meets the constraints of a schema is said to validate against that schema. For example, a schema might constrain its values to arrays of integers between 0 and 10 in which case the value `[2,3,7]` would validate against it while `[0,-2,11]` would not. The main desiderata for a PSI schema language were that it be:

1. Expressive enough to describe natural JSON representations of existing machine learning data structures;
2. Machine interpretable to allow for automated validation;
3. Simple to read and describe.

The *JSON schema language* [6,7,8] is an existing schema language for JSON values which uses JSON objects to describe a variety of constraints on JSON values. It meets the first two requirements above but is arguably too verbose to meet the third requirement. The PSI schema language proposed here mitigates this verbosity through a number of "shorthands" for common JSON schema constraints. This means, like JSON schema, the new schema language can be represented using JSON values, which minimises parsing. However, since the shorthand can be readily translated into a subset of JSON schema, its semantics are grounded by those of JSON schema, which assists validation. Thus, **the PSI framework uses a custom schema language to describe the values that may be passed to or returned from PSI resources.**

An overview of the PSI schema language can be found in Section 1.3. Its details, including of how it is compiled to JSON schema, can be found in Appendix A. The JSON schema specification is divided into three parts:

- core definitions and terminology, available at `http://tools.ietf.org/html/draft-zyp-json-schema-04`;
- validation, available at `http://tools.ietf.org/html/draft-fge-json-schema-validation-00`; and
- hyper-schema, available at `http://tools.ietf.org/html/draft-luff-json-hyper-schema-00`.

## 1.2.5 Composition via References

In order to promote modularity, consistency, and re-use of schema, this specification makes heavy use of, and extends, the reference mechanism of JSON schema. In JSON schema references allow parts of schema to be defined elsewhere and referred to by URI. The references are resolved when needed via a HTTP GET request.

This specification makes two extensions to JSON schema references: the specification of some standard schema that are associated with short names instead of URIs and resolved by the PSI framework; and a simple mechanism for passing arguments to URIs during reference resolution.

**Schema within the PSI framework can refer to other schema via parameterised names and URIs to enable modularity and reuse**.

An overview of schema references and resolution are given in the next section. The details can be found in Appendix A.

## 1.2.6 RESTfulness of the Architecture

Although there are elements of the PSI design that fall short of the REST ideal [1], we believe that the approach is largely RESTful: interaction is via HTTP verbs with well-defined resources; resources exchange representations of themselves and data; and the representations include links to the immediate actions that may be taken by a client and which use a schema language to support customised controls where needed. Two closely-related ways in which PSI is not yet completely RESTful are in the use of a single media type (application/json) for all resource representations and, consequently, a lack of content negotiation. While these may be modified in a future revision of the PSI specification, we have chosen to use JSON for all communication currently because a single, simple and well-known data format eases the implementation path for potential adopters, which in turn supports the ultimate goal of federated machine learning services. JSON is also presentation free: client software developers may provide any suitable presentation of PSI resources and responses using the information contained in the JSON data they receive, including the provision of customised controls based on contained schema.

# 1.3 Values and Schema

**Values** within the PSI framework are used to represent instances and schema when these are required in a request to, or response from, a PSI service.

Values fall into two categories: atomic and structured. *Atomic* values are integers, (real) numbers, strings, or booleans. *Structured* values are arrays or objects. An *array* is an ordered sequence of values (atomic or structured). An *object* is an unordered collection of *properties*: key-value pairs where the key is a always a string and the value can be structured or atomic.

In this specification, values are indicated using a `monospaced font`. Examples of values as defined by the JSON standard include: `11` (integer), `-36.6` (number), `"setosa"` (string), `true` (boolean), `["Ryde", 3, 703.2]` (an array), and `{ "suburb": "Ryde", "bedrooms": 3, "area": 703.2 }` (an object).

A complete specification of the JSON syntax used here is available at `http://json.org`.

A **schema** is a value that describes the structure of other values. Specifically, each schema is an object whose properties describe constraints which other values must satisfy in order to validate against that schema.

A full description of the PSI schema language is given in [Appendix A](#).

An example of a schema that highlights several of these properties is given below:

```
{
    "/version=":   2,
    "/id":         "$integer",
    "/name": {
        "?first":  "$string",
        "/last":   "$string"
    },
    "/addresses": {
        "$array": {
            "items": "$http://example.org/schema/address",
            "minItems": 1
        }
    }
}
```

Each property in this schema defines a constraint determined by the property's key and value. A property key of the form `/KEY` indicates that a value that validates against the schema must be an object and must contain a field `KEY`. A key of the form `?KEY` indicates an optional property. If a value `S` is associated with a key `/KEY` or `?KEY` it indicates that valid values for the schema must have values associated with `KEY` that are valid for the schema `S`. A property with `/KEY=` and value `VALUE` indicates that valid values must have a property with key `KEY` with associated value `VALUE`.

The prefix `$` on a string indicates a schema reference. The references `"$integer"` and `"$string"` are local schema that are standard within the PSI framework to describe integer and string values, respectively. The `"$array"` reference denotes a local schema template. It takes in arguments in the form of a JSON object which controls the schema it returns when resolved. References that are URIs are global schema that must be resolved via HTTP requests. References and their resolution are described in [Section A.2](#) below. Local schema and schema templates are described in [Appendix B](#).

The schema above therefore describes object values with four mandatory fields:

- `version`, which must have the value `2`;
- `id` which must have a value that matches the referenced schema `$integer`;
- `name` which must have an object value with an required field of `last` and optional field `first`, both of which must have values matching the referenced schema `$string`;
- `addresses` which must have a value that is an array with at least one item and with all items matching the referenced schema at `http://example.org/schema/address`.

A value that would match this schema (given a reasonable definition of the address schema at the URI) is:

```
{
    "version": 2,
    "id":      231,
    "name":    { "first": "Amy", "last": "Jones" },
    "address": [ { "number": 14, "street": "Bird St.", "suburb": "Epping" } ]
}
```

### 1.3.1 Rich Values

A large and growing body of machine learning techniques exist for processing complex data such as images, video and audio, which cannot be natively represented in JSON. Such non-JSON values are referred to hereafter as *rich values*. Rich values are represented in PSI requests and responses by strings containing valid URIs, which either point to a resource representing the rich value or directly encode it in the URI string. Rich values may be represented using HTTP URIs or Data URIs, which are distinguished by the URI scheme [9]. HTTP URIs must begin with "http://" while data URIs must begin with "data://".

Rich values are indicated in PSI schema by the prefix @ followed by the media type of the rich value. For example, the PSI schema @image/jpg indicates a rich value that resolves to a JPEG image. An example of a value that would validate against this PSI schema is the string "http://example.org/picture.jpg", assuming that a GET request to that URI returns a response with media type image/jpg.

Validation of rich values depends on the URI scheme used. For a rich value pointed to by an HTTP URI to be valid against the PSI schema @T, a GET or HEAD request to the URI must return a Content-Type field in its response with a value of T. Data URIs, described in RFC2397 [10], embed the media type into the URI itself. For a rich value represented as a Data URI to be valid for the PSI schema @T, the media type in the Data URI must be equal to T.

Full details for handling PSI schema for rich values are given in Section A.3 and A.4.

## 2 The PSI Service Architecture

The main components of the PSI framework are the relation, learner, transformer and predictor resources. As shown in Figure 1, these resources are central to the three primary activities within the framework: *training*, *prediction*, and *updating*. Attributes are required to construct instance representations as JSON values in order to conform to the task schema published by learner resources and may also be used to generate suitable values for transformer and predictor resources. Values for prediction, suitably represented in JSON, may also come from other sources. Transformers may be used to perform one-off calculations or pre-process instances data to assist in learning tasks.

*Training* a learner involves sending it a request with all the information the learner requires to construct a predictor. The totality of this information is called a *task* and typically consists of parameters to configure the learner and one or more representations of instances from a relation. The instance representations are expressed through the use of attributes so as to match the structure the learner requires. The learner expresses its parameter and resource requirements through a *task schema*. As Figure 1 shows the flow of data in the PSI architecture, *task schema* is not explicitly represented in the figure.

To make a *prediction* an instance representation is required to match the structure required by the predictor resource. Once again, this is expressed through the use of schema and may use attributes to construct suitable representations. If a suitable value is given to a predictor as input the predictor will return a *predicted value*.

Some predictor resources can be *updated* after they are initially created. Those that can be updated will provide an *update schema* to express how updating values must be represented before they can be used to update the predictor.

In addition to the main resources involved in the activities of training, prediction and predictor update, there are also resources to assist in exploring the namespace of a PSI service and through which attribute values and predictions are accessed. URIs within resources' representations form the following hierarchy of resources. There is no requirement that a PSI service's namespace reflect this hierarchy—a client or user need only know the URI for the PSI start resource to begin using a service. Further, some resources, such as predictors, may be offered as stand-alone services.

The remainder of this section describes all of the resources within the PSI framework, the ways in which they can be called, and their request and response structures.

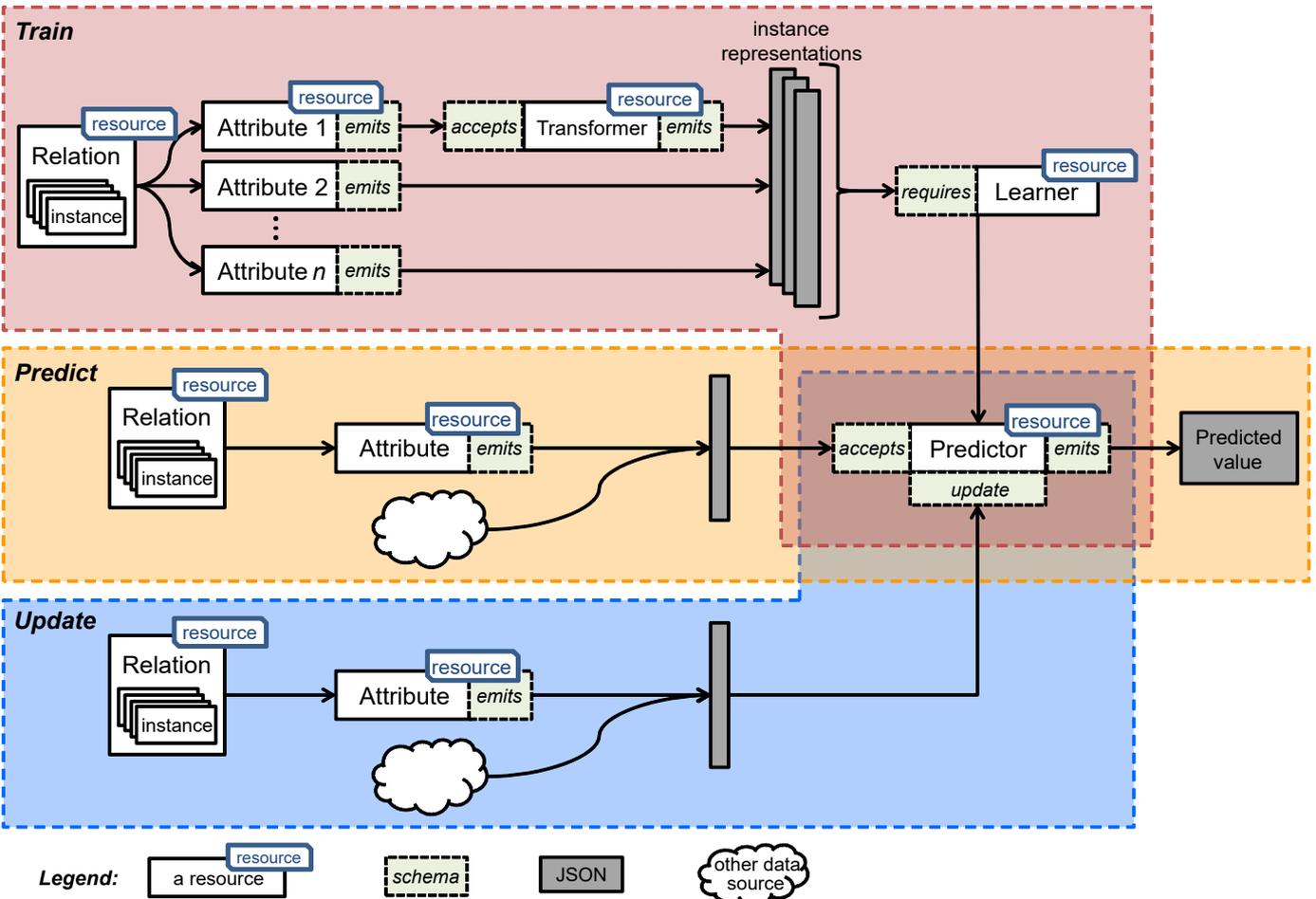

Figure 1: PSI Service Data Flow. In this illustration the output from Attribute 1 is transformed before it is used in learning.

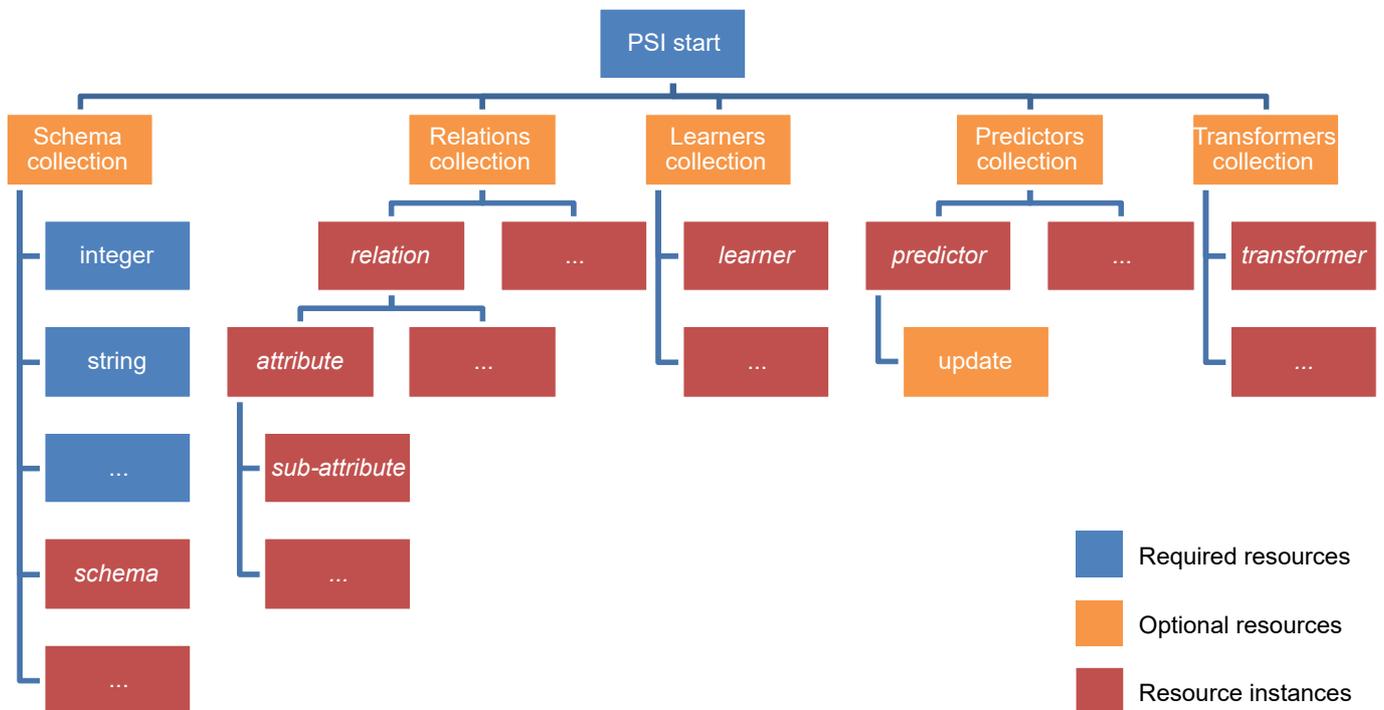

Figure 2: PSI Service Resource Hierarchy. All resource collections are optional and will depend on the nature of the service being offered.

## 2.1 Conventions

Because this specification does not dictate the namespace of a PSI service, URIs in the tables below are represented by short identifiers (such as T or R) that are synonymous with the PSI resource to which the request is sent.

The data sent via POST requests to any PSI resource must be represented in JSON. If a method is described as GET then the arguments passed in via the GET query are assumed to be appropriately URI encoded [9]. Type names refer to those types as represented in JSON. In addition to the standard JSON types, the following pseudo-types, whose names appear in *italics*, are used to represent additional semantics:

- *any* indicates that any JSON value may be used;
- *attribute* indicates a value that represents a new (structured) attribute (defined in Section 2.6.1 below);
- *schema* indicates an object value representing a schema in the PSI schema language; and
- *URI* indicates a string value that can be parsed as a URI (i.e., it can be validated against the `uri` predefined schema given in Section B.1 below).

The string-valued property `psiType` is common to all PSI request and response message bodies. Its value allows servers and clients to determine how to correctly process each message's contents and to validate requests and responses.

## 2.2 Discovering a PSI Service

As each PSI service may use a different namespace for its resources, a mechanism for discovering their URIs must be provided. Each PSI service thus has a single, published entry URI, which can be queried via GET to discover collections of PSI resources available at that service.

| Purpose | Method | URI | Arguments | Returns |
|---|---|---|---|---|
| Begin discovery of a PSI service P | GET | P | None | A description of the PSI service P |

The description of a PSI service contains zero or more properties that are the URIs for collections of the different groups of resources: schema, transformers, relations, learners and predictors (i.e., learned transformers). Each of the collection resources is optional because different services may offer differing subsets of the full service. For instance, one service may only provide data, another may provide only learners and predictors, while another may offer only (pre-trained) predictors.

The description also has an optional `relatedServices` property, which allows different services to provide a set of related actions that may be taken with the information presented in the response. This optional property is an array of Link Description Objects (LDOs) [8] and may appear in any PSI resource representation. Further details, including a description of LDOs, are given in Section 2.2.1 below.

*Service Representation*:

| Property | Type | Required | Description |
|---|---|---|---|
| `psiType` | string | Y | "service" |
| `uri` | URI | Y | URI of the PSI entry point |
| `relations` | URI | N | URI of collection of all relations |
| `schema` | URI | N | URI of collection of all predefined schema |
| `learners` | URI | N | URI of collection of all learners |
| `predictors` | URI | N | URI of collection of all predictors |
| `transformers` | URI | N | URI of collection of all transformers, optionally including predictors |
| `relatedServices` | LDO array | N | List of related actions for this top-level service description |

Each of the five collection resources can report the resources in that collection. Some collections also support creation of additional elements, which is described below where applicable.

| Purpose | Method | URI | Arguments | Returns |
|---|---|---|---|---|
| Discover all resources within a collection C | GET | C | None | A list of URIs for resources in that collection |

The representation of each list of resources is the same, regardless of the kind of resource it contains.

| Property | Type | Required | Description |
| --- | --- | --- | --- |
| `psiType` | string | Y | "resource-list" |
| `uri` | *URI* | Y | URI of the resource collection that produced the list |
| `resources` | *URI* array | Y | List of resources in this collection |
| `relatedServices` | LDO array | N | List of related actions for this resource collection |

### 2.2.1 Discovering Related Services and Actions

While the interactions that a client may have with PSI resources as defined in this specification are sufficient for a range of machine learning activities, some PSI services may wish to offer additional assistance to clients in discovering related actions and resources for each resource with which they interact. For example, a service may wish to indicate which learners may be applied to a given relation. This is afforded by the optional `relatedServices` property, which may be present in any PSI resource representation or response. The property's value is an array of Link Description Objects (LDOs), which are defined as part of JSON Hyper-Schema [8] (but which may be used outside that context). Each LDO must contain an `href` property, which is the target of the link, and a `rel` property describing the relationship of the linked resource to the current resource. There are a number of optional properties for specifying details of the interaction, including the media type of the linked resource, the HTTP method to be used, and the format of acceptable values to send as part of a request, specified in JSON Schema. These are defined in full in the JSON Hyper-Schema specification. As the range of possible interactions is extremely large, the PSI specification does not define any canonical LDOs. Indeed, each LDO should contain sufficient information for a client to determine how to interact with the linked resource.

## 2.3 Schema

Schema are resources within the PSI framework since their representation may be requested during schema resolution as described in Section A.2. The response to a GET request to a schema resource is the JSON value representing that PSI schema.

| Purpose | Method | URI | Arguments | Returns |
| --- | --- | --- | --- | --- |
| Retrieve schema S | GET | S | Schema dependant or `template=true` | The PSI schema for S |

Some schema are *parameterised* in the sense that they can take in arguments and return different variants of the same kind of schema. A description of the parameters a schema takes (if any) can be obtained by setting the `template` argument in the GET request to `true`. Schema requested in this way will return a JSON value called a *schema template* that describes the parameterised values within a schema using strings of the form `%ARG`, where `ARG` denotes an argument name.

For example, a schema for numeric values might have a URI at `http://example.org/schema/number`. When called with no arguments it may return the schema `{ "type": "number" }`. When a GET request to `http://example.org?template=true` is the following schema template is returned:

```
{ "type": "number", "minimum": "%min", "maximum": "%max" }
```

indicating that the schema resource has two arguments `min` and `max` that control the range of values for this schema.

When the same schema resource is called using arguments via a GET to `http://example.org/schema/number?min=10` the schema resource returns `{ "type": "number", "minimum": 10 }`, which constrains valid values to be 10 or larger.

These parameterised schema are used extensively in the predefined schema described in Appendix B. A PSI service must provide the predefined schema defined in Appendix B, but may also provide additional schema.

# 2.4 Transformers

A frequent activity in machine learning is to transform data from one form into another: data may be pre-processed before its use in learning; predictors transform instance data into inferences about that data. Similarly, attributes are (conceptually) transformers that convert some underlying data structure into JSON representations that may be consumed by other resources, although they have different abilities as resources in the PSI framework. A *transformer* resource is essentially a function that maps from JSON values to JSON values. Each transformer has an *accepts schema* that describes the domain of the function it represents and an *emits schema* representing its codomain. Transformers may be joined with other transformers (and with attributes) to create data workflows. A transformer T may be joined with another transformer S (functionally S∘T) if the *emits* schema of T is compatible with the *accepts* schema of S, meaning that any value that is valid for T's *emits* schema must also be value for S's *accepts* schema.

## 2.4.1 Transformers as Resources

A transformer is identified by its URI, to which three types of requests may be issued via HTTP requests: obtain a transformer's representation, apply and join (with another transformer to create a third, new transformer).

### 2.4.1.1 Transformer Representation

| Purpose | Method | URI | Arguments | Returns |
|---|---|---|---|---|
| Describe transformer T | GET | T | none | Description response for T |

The response a transformer resource returns to an GET request with no query contains an identifying response type, the URI of the transformer that the GET request was sent to, and the *accepts* and *emits* schema of the transformer. Optionally, the representation can contain a textual description of the transformer, and an object describing the provenance of the transformer (e.g., when it was created, by whom, etc.).

*Transformer Representation*:

| Property | Type | Required | Description |
|---|---|---|---|
| `psiType` | string | Y | "transformer" |
| `uri` | *URI* | Y | URI of the transformer |
| `description` | string | N | Short, human-readable description of the transformer |
| `accepts` | *schema* | Y | Schema describing valid input values for the transformer |
| `emits` | *schema* | Y | Schema describing the output values of the transformer |
| `provenance` | object | N | A structure describing how this transformer was created |
| `relatedServices` | LDO array | N | List of related actions for this relation |

### 2.4.1.2 Apply

The result of applying a transformer to a value is obtained by sending a GET request to its URI with a `value` argument that is a JSON representation that conforms to the transformer's *accepts* schema. The response contains a JSON value conforming to the transformer's *emits* schema. If the apply request contains a value that does not conform to the *accepts* schema the resource must return HTTP status 400 Bad Request.

| Purpose | Method | URI | Arguments | Returns |
|---|---|---|---|---|
| Apply transformer T to value V | GET | T | `value=V` | A value response |

An apply request has the following property.

*Apply Request*:

| Property | Type | Required | Description |
|---|---|---|---|
| `value` | *any* conforming to *accepts* schema | Y | Input value to transform |

The response to such a request contains the result of applying the transformer to the input value.

*Value Response*:

| Property | Type | Required | Description |
|---|---|---|---|
| psiType | string | Y | "value" |
| value | *any* conforming to *emits* schema | Y | Result of applying the transformer to value in request |
| valueList | *n/a* | N | Absent in transformer responses |
| relatedServices | LDO array | N | List of related actions for this relation |

Messages with `psiType = value` can contain either a `"value"` property that holds a single value or a `"valueList"` property that holds an array of values. As transformers emit only individual values, the `"value"` property is always present in their responses.

### 2.4.1.3 Join

To join a transformer T with another transformer S, a join request is sent via POST to T's URI. The request contains the URI of the other transformer S, whose *accepts* schema is *compatible* with the *emits* schema of T. Compatible here means that any value conforming to the *emits* schema of T must be valid for the *accepts* schema of S. The response to a join request includes the URI for a new transformer S∘T that transforms values by first applying T to the input value and then applying S to the result.

| Purpose | Method | URI | Arguments | Returns |
|---|---|---|---|---|
| Create a new transformer by joining transformer T with transformer S | POST | T | A join request | HTTP status 201 (Created) or 302 (Found). Location header field contains URI of the transformer that joins S and T |

A join request via POST has the form described below.

*Join Request*:

| Property | Type | Required | Description |
|---|---|---|---|
| psiType | string | Y | "composition" |
| join | *URI* | Y | URI for the transformer S to join with this transformer T |
| description | string | N | Human-readable description of the joined transformer |

If successful the service may respond with an HTTP status of 201 (Created) or 302 (Found) and will include the URI of the joined transformer in the Location header field.

## 2.5 Relations and Instances

In the PSI framework, the term *instance* is used to describe a single record of some phenomenon of interest. For example, in the Iris problem, an instance corresponds to a particular flower; in the house ranking problem an instance is a pair of houses ordered by preference; in the document problem an instance is a document. Instances are only ever accessed through their representations and a single instance may support several different representations. For example, a flower might be represented by an array of numbers such as `[ 5.1, 3.5, 1.4, 0.2 ]` or an object like

```
{ "sepal": { "length": 5.1, "width": 3.5 },
  "petal": { "length": 1.4, "width": 0.2 } }
```

Abstractly, a *relation* is a collection of instances which share a common "shape" – that is, they can be represented using the same attributes. Attributes are discussed in more detail below but can be thought of as functions from instances to values. In the Iris example a simple attribute may report the sepal length of an instance while another, more complicated attribute may return an array with all the measurements. Formally then, a relation is a collection of instances that, if one instance is in an attribute's domain, all other instances in that relation are too.

## 2.5.1 Relations as Resources

As a resource, the main function of a relation is descriptive. It must specify how many instances it contains and provide URIs for the attributes that can be used to represent instances. Each instance within a relation is referred to via its unique *index* in the range 1–$n$, where $n$ is the number of instances in the relation.

Both the instance count and default attribute are retrieved as part of a GET request to the relation resource. The GET request may take arguments to select a subset of instances from the relation. These are described in more detail below.

| Purpose | Method | URI | Arguments | Returns |
|---|---|---|---|---|
| Describe relation R | GET | R | (optional) query | Description response for R |

The response a relation resource returns to a GET request contains an identifying response type, the URI of the relation that the GET request was sent to, the size of the relation (i.e., the number of instances, which may vary depending on the query arguments given), the URI of the relation's default attribute, and a list of all attributes for the relation. The default attribute is used to provide some standard way to represent the instances in the relation.

*Relation Representation*:

| Property | Type | Required | Description |
|---|---|---|---|
| `psiType` | string | Y | "relation" |
| `uri` | *URI* | Y | URI of the relation |
| `description` | string | N | A human-readable description of the relation |
| `size` | integer | Y | The number of instances in the relation |
| `defaultAttribute` | *URI* | Y | The default attribute for this relation |
| `attributes` | *URI* array | Y | All attributes defined for this relation |
| `querySchema` | *schema* | N | Schema defining supported query arguments |
| `relatedServices` | LDO array | N | List of related actions for this relation |

## 2.5.2 Selecting Subsets of Instances

A number of machine learning activities require that subsets of instances be used for training and evaluation of predictors. To support this, relation resources can accept query arguments as part of GET requests. In effect, relation resources at URIs `http://example.org/{R}` and `http://example.org/{R}?{query}` represent two different relations that share some instances and may share some or all of their attributes. These differences will be reflected in their representations, which will necessarily have different `uri` properties and will typically have different values for `size`.

If a relation supports queries, its representation will include a `querySchema` property that must be the schema for an object with atomic-valued properties. Each of these properties equates to a GET parameter. If standard schema constraints cannot capture the full complexity of supported queries, then the `description` property of `querySchema` or one of the GET parameters it defines should contain additional, human-readable information, or a URI from which a full description can be obtained.

For example, a PSI service could offer the ability to generate equally-sized subsets of a relation, called *folds*. The query arguments for such a service could be the number of folds $n$ (an integer between 1 and the size of the relation) and the selected fold $i$ (an integer between 1 and the number of folds). For a given number of folds $n$, the entries in fold $i$ could be selected by taking instance $n$, $2n$, $3n$ and so on from the underlying relation.

A relation resource supporting fold selection could accept three GET arguments: the number of folds `numfolds`, the identified `fold` and, optionally, an `invert` argument that inverts the selection. The relation could report the following query schema, which includes schema properties such as `title` and `description` that can be used to "decorate" controls in client software.

```
{
  "description": "Select subset 'fold' of 'numfolds' total subsets of instances.
                  Use 'invert=true' to select every other fold.",
  "/fold":     { "$integer": { "min": 1, "title": "Fold number", "description": "≤ number of folds" } },
  "/numfolds": { "$integer": { "min": 1, "title": "Total folds" } },
  "?invert":   { "$boolean": { "title": "Invert selection"} }
}
```

## 2.6 Attributes

Abstractly, an *attribute* is a function that, when applied to an instance, returns a value. Conceptually, attributes are transformers whose *accepts* schema is hidden from view and whose input values are referred to indirectly via instance indices. In practice, attributes are transformers that accept integers in the range 1 to the size of their relation, as well as the special value `all` to request they transform all instances in their relation, which is a key difference between transformer and attribute resources. Attributes that return atomic values will be referred to as *atomic-valued attributes* and those that return structured values will be referred to as *structure-valued attributes.* Each attribute has an associated *emits* schema. Every value returned by an attribute is valid for that attribute's schema.

Practically, attributes act as the interface between instances and their representation for learning and prediction. Different learning services require different representations of relations in order to extract the information they need to construct predictors. The aim of attributes is to provide a simple, flexible way to build representations of instances that are appropriate for input to a variety of learning and prediction services. Much of this flexibility comes from the ability to compose attributes.

### 2.6.1 Attribute Composition

As with values, arrays and objects can be used to compose attributes to create *structured attributes*. Like structured values, a structured attribute can be created via the recursive use of array and object structures that contain other attributes.

Formally, given $n$ attributes $A_1$, ..., $A_n$, their *array composition* is the attribute that, when given an instance $I$, returns the array value $[V_1, \ldots, V_n]$ where $V_i$ is the value obtained by applying attribute $A_i$ to instance $I$. The *object composition* of the attributes $A_1$, ..., $A_n$ with keywords $K_1$, ..., $K_n$ is the attribute that, when given instance $I$, returns the object value $\{K_1 : V_1, \ldots, K_n : V_n\}$ where the $V_i$ are as before. The recursive application of these two types of composition can be used to construct more complex structured attributes.

The schema associated with the composition of two or more attributes is the corresponding composition of those attributes' schema. If $A$ is the array composition of attributes $A_1$, ..., $A_n$ with schema $S_1$, ..., $S_n$ then the schema for $A$ will be the array composition of the schema $S_1$, ..., $S_n$. If $O$ is the object composition of the same attributes with keys $K_1$, ..., $K_n$ then the schema for $O$ will be the object composition of $S_1$, ..., $S_n$ with the same keys. The details of schema composition are given in [Section A.1.5](#).

For example, suppose $A_1$ and $A_2$ are attributes that return the sepal length and sepal width of an iris and both have the schema `$number`. Furthermore, suppose that the values those attributes return on a particular iris $I$ are 5.1 and 3.5, respectively. The array composition of those attributes would have schema `[ "$number", "$number" ]` and return the array value `[ 5.1, 3.5 ]` when applied to $I$. The object composition of $A_1$ and $A_2$ with keys `length` and `width` would have schema `{ "/length": "$number", "/width": "$number" }` and return the object value `{ "length": 5.1, "width": 3.5 }` when applied to $I$.

### 2.6.2 Attributes as Resources

Like every resource, an attribute is identified by its URI. There are five types of requests that can be made in relation to attributes via HTTP requests: create, obtain an attribute's representation, obtain an attribute's value, join (with a transformer) and delete. Attribute creation is via a request on the new attribute's relation resource. All other requests are sent to the attribute's URI. The format of these requests and their responses are described in the subsections below.

## 2.6.2.1 Create

Since attributes can be composed to form new attributes, it is necessary that there be a way of assigning a new URI to these new attributes. This is the purpose of the create request via the REST POST method. The URI the POST request is sent to is that of the relation for which the new attribute is required. The URI of the new attribute is reported in the response to the create request.

| Purpose | Method | URI | Arguments | Returns |
| --- | --- | --- | --- | --- |
| Create a new attribute A for relation R | POST | R | A create request | HTTP status 201 (Created). Location header field contains new attribute's URI. |

A create request contains a JSON value defining a new attribute to create. This value is either an object (for a new object-composed attribute) or an array (for a new array-composed attribute). Every value in such an object or array is either another value representing an object- or array-composed attribute or a string containing the URI of an existing attribute resource. For example, to create an array composition of the attributes with URIs `http://A1` and `http://A2`, the JSON value to send in the `attribute` field of the create request would be `[ "http://A1", "http://A2" ]`. Each of the existing attributes referred to in the new attribute's definition must belong to the relation to which the create request is submitted.

The create request may also contain a textual description of what the new attribute is for. This is presented in the description field in description responses for the new attribute.

*Create Request*:

| Property | Type | Required | Description |
| --- | --- | --- | --- |
| `psiType` | string | Y | "attribute-definition" |
| `attribute` | *attribute* | Y | Definition of the attribute |
| `description` | string | N | Human-readable description of the new attribute |

If successful the service responds with HTTP status 201 (Created) and includes the new attribute's URI in the Location header field. The schema and other information about the newly created attribute can be obtained via the returned URI's representation (described next).

If the create request was malformed the HTTP client error 400 (Bad Request) is returned. If the server was not able to process the create request for some other reason the HTTP server error 500 (Internal Server Error) is returned.

## 2.6.2.2 Attribute Representation

When queried with a GET request with no arguments, an attribute resource with URI `A` returns a description of itself.

| Purpose | Method | URI | Arguments | Returns |
| --- | --- | --- | --- | --- |
| Describe attribute A | GET | A | None | Description response for A |

The description response contains the URI of the attribute, a short textual description, the schema of values it emits, the URI of its relation and, optionally, its relation's query schema (which may be used to construct query arguments to apply directly to the attribute). If the attribute is composed from other attributes, URIs for these sub-attributes are provided in a structure that corresponds to the attribute's schema (i.e., is either an array or an object). If the attribute is array-valued then sub-attributes are presented as an array of URIs in the same order as their values appear in the attribute's structured value. If the attribute is object-valued then sub-attributes are presented as a collection of name-URI pairs corresponding to the names of properties in the values represented by the attribute.

*Attribute Representation*:

| Property | Type | Required | Description |
| --- | --- | --- | --- |
| psiType | string | Y | "attribute" |
| uri | *URI* | Y | URI of the attribute |
| description | string | N | Short, human-readable description of the attribute |
| emits | *schema* | Y | Schema describing the output of the attribute |
| relation | *URI* | Y | The relation to which the attribute belongs |
| subattributes | array *or* object | N | URIs for sub-attributes if this attribute is structured |
| querySchema | *schema* | N | Schema defining supported query arguments |
| relatedServices | LDO array | N | List of related actions for this attribute |

### 2.6.2.3 Apply

The main function of an attribute is to convert instances and relations into values that can be sent to a learner or predictor resource. The result of applying an attribute to its relation is obtained by sending a GET request to its URI with an `instance` argument that is either the integer index of an instance in the relation (between 1 to the relation's size) or the value `all`. In the first case the attribute will respond with a `value` message, as produced by a transformer applied to some input value. In the second case the attribute will respond with a `value` message with its `"valueList"` property set to an array of values.

| Purpose | Method | URI | Arguments | Returns |
| --- | --- | --- | --- | --- |
| Get value of attribute A for instance *i* or for all instances | GET | A | `instance=`*i* or `instance=all` | Value response for instance *i* or value list response for all instances |

The instances to which the attribute is applied are determined by the information in the request and any query string appended to the attribute's URI. If `instance` is an integer in the appropriate range then the attribute is applied to the instance at that index. If `instance=all` then the attribute is applied to every instance in its relation. If a query string is constructed using the attribute's query schema then these two forms operate on instances in the resultant queried relation (see Section 2.5.2 above).

The response to this request contains the JSON representation of the selected instances. If only a single instance was selected then the response is as as defined for transformers in Section 2.4.1.2 above. If multiple instances were selected then the attribute will respond with a `value` message containing a `"valueList"` property, which is an array of JSON representations.

*Value Response*:

| Property | Type | Required | Description |
| --- | --- | --- | --- |
| psiType | string | Y | "value" |
| value | *any* conforming to *emits* schema | Y* | Value produced by applying the attribute to one instance |
| valueList | array of values conforming to *emits* schema | Y* | Values produced by applying the attribute to all instances |
| relatedServices | LDO array | N | List of related actions for this value response |

Exactly one of `"value"` or `"valueList"` must be included in a message of type `value`.

### 2.6.2.4 Join with a Transformer

A join request for an attribute A contains the URI T of a transformer whose *accepts* schema is compatible with the *emits* schema of A. The request and response behaviour for join requests is identical to that of transformers except that, because the result represents T∘A, the returned URI will correspond to an attribute resource rather than a transformer resource.

### 2.6.2.5 Delete

Client-created attributes may be deleted via a DELETE request sent to the attribute's URI. If the attribute is successfully deleted then the service responds with HTTP status 200 (OK). If the attribute may not be deleted (for instance, it was created by the service's administrators as the default attribute for a relation) then the service responds with HTTP status 403 (Forbidden).

## 2.7 Learners

A learner is a process for generating predictors from relations. Each learner provides a schema for the input it requires to construct a predictor. A value that is valid for a learner's schema is called a *task* and described below. A task typically consists of learning parameters of as well as a relation and attributes to represent instances from the relation in a way that is suitable for the learner. A learner uses the information provided in a task to construct and return a predictor.

### 2.7.1 Tasks and Task Schema

Algorithms that are presented as PSI services have different restrictions on the type of information they can process. For example, an implementation of naive Bayes or a decision tree learner may be able to feature vectors that contain categorical values whereas a support vector machine learner can only accept feature vectors containing real numbers. Similarly, an algorithm for solving a regression problem would require labels for instances to be real numbers while a classifier would need categories for labels. Some learners may additionally require values for parameters that control the learning process. For example, a k-nearest neighbour algorithm requires the number of examined neighbours to be given while a support vector machine may require kernel and regularisation parameters to be set.

The PSI schema language provides a flexible way of specifying the kinds of information a learner requires and the way it is structured. As mentioned above, there are essentially two kinds of values required by a learner: parameters and resources. These two types of value are specified and handled differently.

*Task parameters* can be specified anywhere within a task schema using the PSI schema language. The validation of parameters against those parts of a task schema are handled like any other value validation (see Appendix A). *Task resources* must be specified within a `resources` property within a task schema. Each value within the `resources` property of a task value must be the representation of a PSI resource, i.e., the response from performing a GET request on that resource's URI. These can be provided as URI references, which are URIs with the prefix $. These resource references are dereferenced before their representation values are validated against the resource parts of a task schema. See Sections 5.2 and 5.4 for details of how references are resolved.

As an example, a support vector machine learning algorithm presented as a PSI service for solving classification problems may return a task schema that specifies that the learner requires: an optional regularisation parameter (with a default value of 1), a source attribute for representing instances in the relation as numeric feature vectors, and a target attribute for representing the class labels for instances as `-1` or `1`. Since the target and source attributes are PSI resources, these must be specified within the `resources` property of the task schema. Such a task schema might look like the following:

```
{
  "?lambda":   { "$number": { "min": 0, "default": 1 } },
  "/resources": {
      "/source":   { "$arrayAttribute": { "allItems": "$number" } },
      "/target":   { "$fixedAttribute": { "values": [-1, 1] } }
  }
}
```

Here the schema references for `$number`, `$arrayAttribute`, and `$fixedAttribute` are for predefined schema described in Appendix B. A valid task value for this task schema would be the following:

```
{
  "lambda": 2.3,
  "resources": {
      "source":    "$http://example.org/attribute/iris/features",
      "target":    "$http://example.org/attribute/iris/species"
  }
}
```

This above example assumes that the values returned by GET requests to the resource URIs are valid for the corresponding schema in the `resources` section of the task schema. Further examples of task schema can be found in Section 3 below.

## 2.7.2 Learners as Resources

The overall purpose of learner resource is to take in a task value and return the URI of a trained predictor resource for that task. A *describe* request can be made via GET to a learner resource to determine the structure of the task the learner requires. Once a suitable task is assembled a *process* request can be sent via POST to the learner to construct a predictor resource and return its URI.

### 2.7.2.1 Learner Representation

A description of a learner resource is obtained by sending a request via GET to the learner's URI. No arguments are required for this request.

| Purpose | Method | URI | Arguments | Returns |
| --- | --- | --- | --- | --- |
| Describe learner L | GET | L | None | A description response for L |

The response to a description request contains the URI of the learner being described, some text describing the learner, and a schema for what is required in the `task` field of a process request to this learner.

*Learner Representation*:

| Property | Type | Required | Description |
| --- | --- | --- | --- |
| `psiType` | string | Y | "learner" |
| `uri` | *URI* | Y | URI of the learner being described |
| `description` | string | Y | Short, human-readable description of learner |
| `taskSchema` | *schema* | Y | Schema specifying format of tasks this learner can process |
| `relatedServices` | LDO array | N | List of related actions for this learner |

### 2.7.2.2 Process

A process request sent via POST to a learner resource is used to start learner training on the task specified in the process request.

| Purpose | Method | URI | Arguments | Returns |
| --- | --- | --- | --- | --- |
| Apply learner L to a task T | POST | L | A learner process request describing T | HTTP status of 201 (Created) 202 (Accepted). Location header field contains new predictor's URI. |

The process request must contain a JSON value representation of a task, which must be valid for the schema returned by the learner in the `taskSchema` field of its description response.

*Process Request*:

| Property | Type | Required | Description |
| --- | --- | --- | --- |
| psiType | string | Y | "task" |
| task | object | Y | Definition of the learning task to be processed; must conform to the schema defined in the taskSchema property of a learner's description |

If training a predictor completes quickly the learner resource may respond immediately to a process request with HTTP status 201 (Created) and the location of the new predictor resource. Since training a predictor can take hours or even days, a learner resource may also respond with HTTP status 202 (Accepted). In this case, the Location header field still points to a predictor resource, but until training completes that resource will return an alternative representation that describes the current state of training.

## 2.8 Predictors

A predictor is a transformer that is constructed by a learner. It thus has the same representation and behaviour as any other transformer, and so may be composed with other transformers or with an attribute (to provide predictions over a relation). Predictors may differ in four ways from other transformers: their provenance may link to the learner that created them; they have a different representation if they are still being trained; they may support additional training; and they may be deleted.

If a predictor is the result of processing a learning task then its provenance field must contain a `learner` property whose value is the URI of the learner that produced it. Since a "predictor", in the sense of a transformer that makes inferences about data, can conceivably be hand-crafted, this property may be absent.

If a predictor resource is still being trained then a GET request to its URI returns an alternative *status* representation with the following properties. Any other requests will result in HTTP status 403 (Forbidden) until the predictor has been trained, after which the predictor's representation will be as described above.

*Status Response*:

| Property | Type | Required | Description |
| --- | --- | --- | --- |
| psiType | string | Y | "training-status" |
| uri | URI | Y | URI of the predictor |
| learner | URI | Y | URI of the learner processing the learning task |
| status | string | Y | Text describing the status of the learning process |
| relatedServices | LDO array | N | List of related actions for this status |

### 2.8.1 Updating Predictors

Some predictors support being modified in response to additional training examples. Such predictors are called *updatable predictors* and expose an additional interface through which these extra training examples can be given. Additional training examples are referred to as *update values*. The URI for the update interface is given by an `update` property in the predictor's representation, which augments the properties present for other transformers. The required format of an update value is obtained via a GET request on the update resource.

*Obtain Update Schema Request*:

| Purpose | Method | URI | Arguments | Returns |
| --- | --- | --- | --- | --- |
| Discover predictor P's update request schema from its update resource U | GET | U | None | Schema defining the structure of update values used in update requests |

The response to this request is a PSI schema defining the structure of update values. One or more update values may be enclosed in a `value` message and sent via POST to the update interface's URI.

| Purpose | Method | URI | Arguments | Returns |
|---|---|---|---|---|
| Update predictor P using its update resource U | POST | U | A `value` message | HTTP status of 201 (Created) or 303 (See Other) with location of updated predictor |

The request message body for an update has the same structure as `value` messages emitted by transformers and attributes, but with the following restrictions on property values.

*Update Request*:

| Property | Type | Required | Description |
|---|---|---|---|
| `psiType` | string | Y | "value" |
| `value` | *any* conforming to update schema | Y* | A single update value, which is valid against the predictor's update schema |
| `valueList` | array of values conforming to update schema | Y* | An array of update values, all of which are valid against the predictor's update schema |

Exactly one of `"value"` and `"valueList"` must be included. If the request was processed successfully then the service may respond with either 201 (Created) or 303 (See Other), with the Location header field set to the modified predictor's URI. In the first case the location must be different from the original predictor's, and implies that predictor resources at the service in question are immutable. In the second case the URI may be the same as the original predictor's.

### 2.8.2 Deleting a Predictor

A predictor may be deleted by sending a DELETE request to its URI. If successful then the service responds with HTTP status 200 (OK).

## 2.9 Minimum Requirements to be a PSI Service

A full PSI service would consist of all the resource types described above. However, individual PSI resources can exist in isolation and still constitute a "service". For example, a previously-trained predictor resource could be made available for public use, without the provider also hosting the training relation and learning algorithm. The following table summarises the expected resources in services offering different subsets of PSI resources.

| Service includes | Service *must* also have | Service *should* also have |
|---|---|---|
| 1 relation | attributes | |
| many relations | attributes | relation collection, main entry point |
| 1 learner | predictors, status resources | predictor collection, main entry point, learner collection (with one entry) |
| many learners | predictors, status resources | learner collection, predictor collection, main entry point |
| 1 predictor | | |
| many predictors | | predictor collection, main entry point |
| 1 transformer | | |
| many transformers | | transformers collection, main entry point |
| custom schema | schema collection | main entry point |

# 3 Example Usage

This section presents how a hypothetical set of PSI services might be interacted with. This includes: discovering the service and its resources; examining a relation consisting of iris instances using its attributes; examining a learner to determine its task schema; creating a new attribute for the relation in order to construct an appropriate task for the learner; examining the resultant predictor; predicting with the predictor; updating the predictor; joining the instance-representing attribute of a different relation to the predictor to make predictions.

The example calls to PSI resources are presented using the following schema, where each part is enclosed in angle brackets < >:

```
<HTTP method> <Example resource's URI>
<JSON encoded request object, if applicable>
----
<HTTP response code and label>
<HTTP headers, if applicable>
<JSON encoded response object, if applicable>
```

## 3.1 Discovering a PSI service

A new PSI service is made available and has a published entry point at the URI `http://example.org`. To discover the namespace of this service and the resources it offers a GET request can be sent to that URI.

```
GET http://example.org
----
200 OK
{
    "psiType":         "service",
    "uri":             "http://example.org",
    "relations":       "http://example.org/data",
    "schema":          "http://example.org/schema",
    "learners":        "http://example.org/learn",
    "predictors":      "http://example.org/infer",
    "transformers":    "http://example.org/transform",
    "relatedServices": [
        { "rel": "help", "href": "http://psi.cecs.anu.edu.au/spec" },
    ]
}
```

This top-level service description also includes a related service representing help information in the form of this specification document. Link description objects can describe far more complex interactions, but no further examples are given in this section as the `relatedServices` field is made available for diverse service providers to describe interactions outside this specification.

The resource collections for transformers, learners and predictors will be examined as needed in later examples. However, to begin a user may be interested in what data sets are available at the service, which can be obtained using GET on the `relations` collection URI.

```
GET http://example.org/data
----
200 OK
{
    "psiType":     "resource-list",
    "uri":         "http://example.org/data",
    "resources":   [ "http://example.org/data/iris" ]
}
```

In this example the service offers only a single relation representing the Iris data set, the details of which can be inspected next.

## 3.2 Examining a relation using its default attribute

A version of the Iris data set is made available as a PSI relation at the URI `http://example.org/data/iris`, which was obtained from the service's relations collection. A description of the relation is obtained using a GET request to that URI.

```
GET http://example.org/data/iris
----
200 OK
{
    "psiType":         "relation",
    "uri":             "http://example.org/data/iris",
    "description:      "The iris data set, courtesy of Sir R. A. Fisher",
    "size":            150,
    "defaultAttribute": "http://example.org/data/iris/flower",
    "attributes":      [
        "http://example.org/data/iris/flower",
        "http://example.org/data/iris/features",
        "http://example.org/data/iris/image
    ],
    "querySchema": { ... }
}
```

This description response shows that the relation has 150 instances and they can be represented using the default attribute at `http://example.org/data/iris/flower`. The relation has two additional attributes at `http://example.org/data/iris/features` and `http://example.org/data/iris/image`. The relation's query schema has been omitted for brevity, but is examined in Section 3.2.1 below. A GET request to the default attribute URI reveals its description, *emits* schema, relation, and sub-attributes:

```
GET http://example.org/data/iris/flower
----
200 OK
{
    "psiType":     "attribute",
    "uri":         "http://example.org/data/iris/flower",
    "description": "A structured attribute for presenting iris dimensions",
    "relation":    "http://example.org/data/iris",
    "emits":       {
        "/sepal": { "/length": "$number", "/width": "$number" },
        "/petal": { "/length": "$number", "/width": "$number" }
        "/species": { "$string": { "enum": [ "setosa", "versicolor", "virginica" ] } }
    },
    "subattributes": {
        "sepal":   "http://example.org/data/iris/flower/sepal",
        "petal":   "http://example.org/data/iris/flower/petal",
        "species": "http://example.org/data/iris/flower/species"
    },
    "querySchema": { ... }
}
```

From this description it is clear that this attribute returns values that are objects with a `sepal` and `petal` properties, and the value for those properties are objects with `length` and `width` properties that must contain numeric values. It also shows that this attribute returns values with a `species` property that can take on the string values "setosa", "versicolor" or "virginica".

The description also shows that this structured attribute has three sub-attributes, accessible at `http://example.org/data/iris/flower/sepal`, `http://example.org/data/iris/flower/petal` and `http://example.org/data/iris/flower/species`. GET requests to these URIs can be used to obtain more information. For example, the `petal` sub-attribute responds with:

```
GET http://example.org/data/iris/flower/petal
----
200 OK
{
    "psiType":      "attribute",
    "uri":          "http://example.org/data/iris/flower/petal",
    "relation":     "http://example.org/data/iris",
    "emits":        { "/length": "$number", "/width": "$number" },
    "subattributes": {
        "length": "http://example.org/data/iris/flower/petal/length",
        "width": "http://example.org/data/iris/flower/petal/width"
    },
    "querySchema": { ... }
}
```

To get the value of the first instance in the iris relation using the default attribute the following GET request can be made:

```
GET http://example.org/data/iris/flower?instance=1
----
200 OK
{
    "psiType":  "value",
    "value":    {
        "sepal": { "length": 5.1, "width": 3.5 },
        "petal": { "length": 1.4, "width": 0.2 },
        "species": "setosa"
    }
}
```

Applying the `species` sub-attribute to the entire Iris relation yields an array of string values (the ellipsis denotes 145 elided values in the array):

```
GET http://example.org/data/iris/flower/species?instance=all
----
200 OK
{
    "psiType":   "value",
    "valueList": [
        "setosa", "setosa", "setosa", ..., "virginica", "virginica"
    ]
}
```

### 3.2.1 Examining a derived relation

This particular PSI service offers the ability to select folds in a data set in the manner described in Section 2.5.2 above. The form of these queries is given by the `querySchema` property in the relation and its attributes. Below is the result of getting the relation's representation (as above), but with all properties other than `querySchema` hidden:

```
GET http://example.org/data/iris
----
200 OK
{
    ...
    "querySchema": {
        "description": "Select subset 'fold' of 'numfolds' total subsets of instances.
                        Use 'invert=true' to select every other fold.",
        "/fold":     { "$integer": { "min": 1, "title": "Fold number", "description": "≤ number of folds" } },
        "/numfolds": { "$integer": { "min": 1, "title": "Total folds" } },
        "?invert":   { "$boolean": { "title": "Invert selection"} }
    }
}
```

This schema includes information about valid values for the query arguments `fold`, `numfolds` and `invert` as well as field titles and descriptions that could be used to "decorate" user controls in client software. Based on this schema, the second of five folds can be requested:

```
GET http://example.org/data/iris?fold=2&numfolds=5
----
200 OK
{
    "psiType":          "relation",
    "uri":              "http://example.org/data/iris",
    "description":      "The iris data set, courtesy of Sir R. A. Fisher (fold 2 of 5)",
    "size":             30,
    "defaultAttribute": "http://example.org/data/iris/flower?fold=2&numfolds=5",
    "attributes":       [
        "http://example.org/data/iris/flower?fold=2&numfolds=5",
        "http://example.org/data/iris/features?fold=2&numfolds=5"
        "http://example.org/data/iris/image?fold=2&numfolds=5"
    ]
    "querySchema": { ... }
}
```

Note that this derived relation contains only 30 instances and reports a different URI for its attributes (the same query schema can be used to construct these URIs directly). Requesting the value of the queried default attribute applied to instance 1 produces the following:

```
GET http://example.org/data/iris/flower?fold=2&numfolds=5&instance=1
----
200 OK
{
    "psiType":  "value",
    "value":    {
        "sepal": { "length": 4.9, "width": 3.0 },
        "petal": { "length": 1.4, "width": 0.2 },
        "species": "setosa"
    }
}
```

Because of the way the folds are selected, instance 1 from this relation is not the same as that from the original relation at `http://example.org/data/iris`.

## 3.2.2 Transforming an attribute's value

Consider the case where a user who wants to manipulate an attribute's value before using it in training. To begin, the user can request the list of transformers available at the service:

```
GET http://example.org/transform
----
200 OK
{
    "psiType":      "resource-list",
    "uri":          "http://example.org/transform",
    "resources":    [
        "http://example.org/transform/square",
        "http://example.org/transform/average"
    ]
}
```

This indicates there are two transformers available. Requesting a representation of the first produces the following:

```
GET http://example.org/transform/square
----
200 OK
{
    "psiType":      "transformer",
    "uri":          "http://example.org/transform/square",
    "description":  "Calculates the square of a number",
    "accepts":      "$number",
    "emits":        "$number",
    "provenance": {
        "created":   "2013-08-15T09:00Z",
        "createdBy": "system"
    }
}
```

This response indicates that the transformer calculates the squares of real-valued numbers. This behaviour can be tested by requesting the transformer's value for the input 4:

```
GET http://example.org/transform/square?value=4
----
200 OK
{
    "psiType": "value",
    "value":   16.0
}
```

To transform all of the values of the Iris relation's attribute for sepal length, a join request must first be sent via POST to that attribute's URI:

```
POST http://example.org/data/iris/flower/sepal/length
{
    "psiType": "composition",
    "join":    "http://example.org/transform/square"
}
----
201 Created
Location: http://example.org/data/iris/flower/sepal/length?t=W1sidC9zcXVhcmUiLCIkbnVtYmVyIl1d
```

This example service encodes transformer joins as a URI query argument, but any naming scheme is allowed. A normal request to apply the transformed attribute to all instances is then made (the ellipsis indicates omitted values):

```
GET http://example.org/data/iris/flower/sepal/length?t=W1sidC9zcXVhcmUiLCIkbnVtYmVyIl1d&instance=all
----
200 OK
{
    "psiType":      "value",
    "valueList":    [
        26.01, 24.01, 22.09, ..., 38.44, 34.81
    ]
}
```

This new, transformed attribute could then be used in learning. However, the following examples consider the case where the untransformed attribute values are used instead. An example in Section 3.9 below shows how this same process can be used to create an attribute that makes predictions about the data in its relation.

## 3.3 Examining a learner

A GET request to the `learners` collection URI returns the following list of available learner resources:

```
GET http://example.org/learn
----
200 OK
{
    "psiType":      "resource-list",
    "resources":    [
        "http://example.org/learn/c45",
        "http://example.org/learn/imageclass",
        "http://example.org/learn/knn",
        "http://example.org/learn/naivebayes"
    ]
}
```

The learner available at `http://example.org/learn/knn` is a simple k-nearest neighbour learning algorithm. A GET request to this URI gives the following response:

```
GET http://example.org/learn/knn
----
200 OK
{
    "psiType":      "learner",
    "uri":          "http://example.org/learn/knn",
    "description":  "A k-nearest neighbour algorithm that takes feature vectors as input",
    "taskSchema":   {
        "?k": { "$integer": { "default": 1, "min": 1,
                    "description": "The number of nearest neighbours to examine" } },
        "/resources": {
            "/target":   { "$nominalAttribute": { "allItems": "$string" },
            "/source":   { "$arrayAttribute": { "allItems": "$atomicValueSchema" } }
        }
    }
}
```

The above description shows (with reference to the predefined schema in Appendix B) that the learner requires a target attribute that returns string values to be interpreted as nominal values, and a source attribute that returns arrays of atomic values (i.e., feature vectors). The learner can also take an optional parameter `k` which controls the number of nearest neighbours to consider. By default, this is set to 1.

# 3.4 Constructing a task

To build a task for the knn learner that will learn to classify irises, the information provided by the default attribute for the iris relation needs to be re-organised to present the iris dimensions as an array of numbers. This can be done by creating a new attribute using a create request sent via POST to the iris relation:

```
POST http://example.org/data/iris
{
    "psiType":     "attribute-definition",
    "description": "A feature vector representation of iris dimensions",
    "attribute": [
        "http://example.org/data/iris/flower/sepal/length",
        "http://example.org/data/iris/flower/sepal/width",
        "http://example.org/data/iris/flower/petal/length",
        "http://example.org/data/iris/flower/petal/width"
    ]
}
----
201 Created
Location: http://example.org/data/iris/array1
```

A GET request to the new URI reveals the schema of values emitted by the newly created attribute:

```
GET http://example.org/data/iris/array1
----
200 OK
{
    "psiType":     "attribute",
    "uri":         "http://example.org/data/iris/array1",
    "description": "A feature vector representation of iris dimensions",
    "relation":    "http://example.org/data/iris",
    "emits":       { "$array": { "items": [ "$number", "$number", "$number", "$number" ] } },
    "subattributes": [
        "http://example.org/data/iris/flower/sepal/length",
        "http://example.org/data/iris/flower/sepal/width",
        "http://example.org/data/iris/flower/petal/length",
        "http://example.org/data/iris/flower/petal/width"
    ],
    "querySchema": { ... }
}
```

The above response shows that the newly created attribute emits arrays of four real-valued numbers. The list of sub-attributes refers to the existing attributes used to define the new attribute in its creation request.

## 3.5 Training a predictor

Since the newly created attribute returns arrays of numbers, it is valid for the source part of the task schema in the knn learner's description shown above. The following process request can be sent to the knn learner via POST:

```
POST http://example.org/learn/knn
{
    "psiType": "task",
    "task": {
        "k": 3,
        "resources": {
            "source":   "$http://example.org/data/iris/array1",
            "target":   "$http://example.org/data/iris/flower/species"
        }
    }
}
----
201 Created
Location: http://example.org/infer/knn_iris_20130802180742606
```

K-nearest neighbour algorithms are "lazy" algorithms in the sense that training consists of merely memorising the instances. This means the training time for this task is very short and so the learner is able to return the HTTP status code 201 (Created) and the URI for the predictor immediately, rather than the HTTP status 202 (Accepted).

## 3.6 Examining a predictor

A GET request to the URI of the new predictor results in a description response:

```
GET http://example.org/infer/knn_iris_20130802180742606
----
200 OK
{
    "psiType":      "transformer",
    "uri":          "http://example.org/infer/knn_iris_20130802180742606",
    "description":  "kNN trained predictor (trained on iris)",
    "accepts":      { "$array": { "items": [ "$number", "$number", "$number", "$number" ] } },
    "emits":        { "$string" { "enum": [ "setosa", "versicolor", "virginica" ] } },
    "provenance":   {
        "learner": "http://example.org/learn/knn",
        "task":    {
            "k":           3,
            "resources": {
                "source":   "$http://example.org/data/iris/array1",
                "target":   "$http://example.org/data/iris/flower/species"
            }
        },
        "created": "2013-08-02T08:07Z"
    },
    "update":       "http://example.org/infer/knn_iris_20130802180742606/update"
}
```

This response gives a lot of detail about the predictor and how to use it. The optional `provenance` property contains an object that gives the learner URI and task that was used to create this predictor. The `accepts` schema defines valid input values, which in this case is arrays of four real-valued numbers. The `emits` schema shows that this predictor's output values are one of three possible string values: "setosa", "versicolor" or "virginica". The presence of the `update` URI indicates this predictor can be updated with additional training instances.

## 3.7 Using a predictor

To use this predictor to classify a previously unseen instance value is identical to applying a transformer, so the input value must be valid for the `accepts` schema. In this case `accepts` defines arrays of four real-valued numbers, so the value `[ 6.1, 2.1, 4.1, 1.7 ]` would be an appropriate one for which to request a prediction:

```
GET http://example.org/infer/knn_iris_20130802180742606?value=[6.1,2.1,4.1,1.7]
----
200 OK
{
    "psiType":      "value",
    "value":        "versicolor"
}
```

## 3.8 Updating a predictor

If a new iris was found, manually classified, and the appropriate measurements made, the predictor created above could be updated to take into account this instance. To do so the instance and its classification would need to be represented as a value that is valid for the predictor's update schema, which can be obtained from the predictor's update resource via GET:

```
GET http://example.org/infer/knn_iris_20130802180742606/update
----
200 OK
{
    "/target":  { "$string" { "enum": [ "setosa", "versicolor", "virginica" ] } },
    "/source":  {
        "$array": { "items": [ "$number", "$number", "$number", "$number" ] }
    }
}
```

If the new iris was determined to be a "virginica" with sepal length of 6.4, a sepal width of 3.1, a petal length of 6.5, and a petal width of 2.1, this information would be presented to the predictor in the following update request:

```
POST http://example.org/infer/knn_iris_20130802180742606/update
{
    "psiType":      "value",
    "value": {
        "target":   "virginica",
        "source":   [ 6.4, 3.1, 6.5, 2.1 ]
    }
}
----
303 See Other
Location: http://example.org/infer/knn_iris_20130802180742606
```

The response returns the URI for the updated predictor in the HTTP headers which, in this case, is the same as the original predictor's URI. A description request to this URI shows how the predictor has been updated:

```
GET http://example.org/infer/knn_iris_20130802180742606
----
200 OK
{
    "psiType":       "transformer",
    "uri":           "http://example.org/infer/knn_iris_20130802180742606",
    "description":   "kNN trained predictor (trained on iris)",
    "accepts":       { "$array": { "items": [ "$number", "$number", "$number", "$number" ] } },
    "emits":         { "$string" { "enum": [ "setosa", "versicolor", "virginica" ] } },
    "provenance":    {
        "learner":   "http://example.org/learn/knn",
        "task":      {
            "k":           3,
            "resources": {
                "source":   "$http://example.org/data/iris/array1",
                "target":   "$http://example.org/data/iris/flower/species"
            }
        },
        "created":   "2013-08-02T08:07Z",
        "updated":   "2013-08-15T12:06Z"
    },
    "update":        "http://example.org/infer/knn_iris_20130802180742606/update"
}
```

In this case, the only thing that has changed is the addition of the `updated` property to the `provenance` part of the response. This is not a requirement of the PSI specification and other predictors may choose to represent changes due to updates in a different manner.

## 3.9 Predicting on a new relation

One important use of trained predictors is as attributes for new relations. Suppose an online flower retailer wanted to automatically classify irises it had in its stock database using the predictor constructed above. It could do so by joining one of its data set's attributes to the predictor, provided that the attribute *emits* values that the predictor *accepts*.

The online flower retailer's database has sepal and petal measurements for its irises but these are accessed through the following attribute (note that, unlike earlier examples, this domain exists, but does not host a PSI service):

```
GET http://flowers.com/data/irises/dimensions
----
200 OK
{
    "psiType":       "attribute",
    "uri":           "http://flowers.com/data/irises/dimensions",
    "description":   "Petal and sepal dimensions as [ sepal length, sepal width, petal length, petal width ]",
    "relation":      "http://flowers.com/data/irises",
    "emits":    {
        "$array": { "items": [ "$number", "$number", "$number", "$number" ] }
    },
    "subattributes": [
        "http://flowers.com/data/irises/dimensions/1",
        "http://flowers.com/data/irises/dimensions/2",
        "http://flowers.com/data/irises/dimensions/3",
        "http://flowers.com/data/irises/dimensions/4"
    ]
}
```

This attribute is conveniently already compatible with the values that the predictor accepts. The following join request connects the attribute to the knn predictor at the example.org service:

```
POST http://flowers.com/data/irises/vector
{
    "psiType":      "composition",
    "join":         "http://example.org/infer/knn_iris_20130802180742606",
    "description":  "Predicted species"
}
----
201 OK
Location: http://flowers.com/irises/example.org/infer/knn_iris_20130802180742606
```

The new attribute can predict the species of irises in the retailer's database:

```
GET http://flowers.com/data/irises/example.org/infer/knn_iris_20130802180742606
----
200 OK
{
    "psiType":      "attribute",
    "uri":          "http://flowers.com/data/irises/example.org/infer/knn_iris_20130802180742606",
    "description":  "Predicted species",
    "relation":     "http://flowers.com/data/irises"
    "emits":        { "$string" { "enum": [ "setosa", "versicolor", "virginica" ] } },
}
```

## 3.10 Training and prediction on rich values

Consider if Sir Ronald Fisher had taken colour photographs of the 150 irises represented in the Iris relation, and that those images were accessible in this example service via the attribute at `http://example.org/data/iris/image`. Inspecting it reveals that it emits JPEG images, either referred to via "http" URIs or encoded in "data" URIs:

```
GET http://example.org/data/iris/image
----
200 OK
{
    "psiType":      "attribute",
    "uri":          "http://example.org/data/iris/image",
    "relation":     "http://example.org/data/iris",
    "emits":        "@image/jpeg",
    "querySchema":  { ... }
}
```

Requesting the image of the first instance in the Iris relation at `http://example.org` produces a data URI-encoded JPEG image (in which the ellipsis indicates the 26,500 character URI has been truncated):

```
GET http://example.org/data/iris/image?instance=1
----
200 OK
{
    "psiType":      "value",
    "value":        "data:image/jpeg;base64,/9j/4AAQSkZJRgABAQEA..."
}
```

The list of learners retrieved earlier included the following:

```
GET http://example.org/learn/imageclass
----
200 OK
{
    "psiType":      "learner",
    "uri":          "http://example.org/learn/imageclass",
    "description":  "Supervised classifier of JPEG images using colour and shape information",
    "taskSchema":   {
        "/resources": {
            "/target":   { "$nominalAttribute": { "allItems": "$string" },
            "/source":   { "$richValueAttribute": { "mediaType": "image/jpeg" } }
        }
    }
}
```

As the image-producing attribute emits rich values that are JPEG images it is suitable for use in training with the imageclass classifier. Training a predictor proceeds in the same manner as the earlier example. Examining the produced predictor:

```
GET http://example.org/infer/imageclass_iris_20130816100025432
---
200 OK
{
    "psiType":      "transformer",
    "uri":          "http://example.org/infer/imageclass_iris_20130816100025432",
    "description":  "Image classifier (trained on iris)",
    "accepts":      "@image/jpeg",
    "emits":        { "$string" { "enum": [ "setosa", "versicolor", "virginica" ] } },
    "provenance":   {
        "learner": "http://example.org/learn/imageclass",
        "task":    {
            "resources": {
                "source":   "$http://example.org/data/iris/image",
                "target":   "$http://example.org/data/iris/flower/species"
            }
        },
        "created": "2013-08-16T0000Z"
    }
}
```

Prediction on JPEG images of iris flowers is now possible, either by providing the encoded image directly in a data URI or indirectly via its HTTP URI:

```
GET http://example.org/infer/imageclass_iris_201308160900432?
value=http://upload.wikimedia.org/wikipedia/commons/5/56/Kosaciec_szczecinkowaty_Iris_setosa.jpg
----
200 OK
{
    "psiType": "value",
    "value":   "setosa"
}
```

# 4 Future Revisions

The following will be the subject of future revisions of the PSI specification:

- Allowing schema resources to validate values via POST.
- The optional use of compressed JSON representations for large instance representations.
- Security, encryption and access control.

---

# Appendix A: The PSI Schema Language

This appendix provides technical definitions of the PSI schema language and details of how it is compiled down to the JSON schema language.

## A.1 Schema

A **schema** is a JSON value that describes the structure of other JSON values. Specifically, each schema is an object composed of zero or more properties with predefined keys. Each of a schema's properties defines a constraint on the values that are valid for that schema. Specifically, we will say a value V is *valid* for a schema S if V satisfies all the constraints represented by properties in S.

An important feature of schema is that, since schema are also JSON values, their structure can be described and validated by a "meta-schema" that is handled in exactly the same way as a normal schema.

The following constraints are a subset of those defined by JSON Schema [7]:

### A.1.1 Type and Range Constraints

The following constraint restricts the type of JSON value that is considered valid. The variable T denotes either a string describing a JSON type (e.g., integer, number, string, boolean, array, object), or is an array of such strings, denoting a disjunction of types.

`"type": T`
: The value T in this *type* constraint can be one of the following six strings: `"integer"`, `"number"`, `"string"`, `"boolean"`, `"array"`, or `"object"`; or T can be an array value containing a subset of those strings. If T is a string value, the constraint is only satisfied by values V corresponding to T (e.g., if T is '"integer"' then V must be an integer, etc.). If T is a list value then the constraint is only satisfied for values corresponding to one of the type in that list (e.g., if T is `["integer", "string"]` then V must be an integer *or* a string).

The next two constraints can be used to limit the allowable range of a numeric value (i.e., a JSON number or integer). The variable N denotes a value that is a number or an integer.

`"minimum": N`
: This *minimum value* constraint is satisfied by any numeric value greater than or equal to N.

`"maximum": N`
: This *maximum value* constraint is satisfied by any numeric value less than or equal to N.

The next constraint is only satisfied by a finite number of specific values. The variable A denotes the array of values that can be matched.

`"enum": A`
: This *enumeration* constraint is satisfied by any value V that appears in the array A.

The following constraints ensure that a value is valid against one or more additional schema. The array A contains the additional schema to be used in validation. It can be used to "import" the constraints from one or more schema into another.

`"allOf": A`

    This schema constraint is satisfied by a value that is valid against every schema in A.

`"oneOf": A`

    This schema constraint is satisfied by a value that is valid against exactly one schema in A.

The JSON Schema validation reference [7] also defines `anyOf` and `noneOf` constraints, but they are not used in the predefined schema described below.

## A.1.2 Array Constraints

The following constraints restrict the size and contents of a JSON array. The variable A denotes an array of schema; S denotes a schema; and N denotes an integer value.

`"minItems": N`

    This *minimum items* constraint is satisfied by any array value with at least N items.

`"maxItems": N`

    This *maximum items* constraint is satisfied by any array value with at most N items.

`"allItems": S`

    This *all items* constraint is satisfied by any array value whose items are all valid for the schema S.

`"items": A`

    This *items* constraint is satisfied by any array value V with the same number of items as the array A and each item in V is valid for the schema in A at the corresponding position.

## A.1.3 Object Constraints

In the following constraint definitions the variable K denotes a keyword that can be JSON strings except those with prefix `/` or `?` or suffix `=` or `*`; the variable S denotes some arbitrary schema; and C denotes any JSON value.

`"/K": S`

    This *mandatory property* constraint is satisfied by any value V that is an object containing a key K with associated value A such that A is valid for the schema S.

`"?K": S`

    This *optional property* constraint is satisfied by any value V that is an object and does not contain the key K. It is also satisfied by any value V that is an object with key K provided the its associated value A such that A is valid for the schema S.

`"/K=": C`

    This *mandatory property value* constraint is satisfied by any value V that is an object containing the key K with associated value equal to C.

`"?K=": C`

    This *optional property value* constraint is satisfied by any value V that is an object and does not contain the key K. It is also satisfied by any value V with key K provided its associated value equal to C.

`"/*": S`

    This *additional properties* constraint will only validate a value V that is an object such that each property in V that is not matched by another constraint has a value that is valid for the schema S.

## A.1.4 Locally Defined Schema

In order to avoid long schema having to be written out twice, the PSI schema language allows names to temporarily associated with schema in the resolution and compilation processes described below. A locally defined schema is indicated by a property key that begins with the `#` character.

`"#K": S`

    This property associates a name K with the schema S. References of the form `"$K"` elsewhere in the same object as the one containing the property are resolved to the schema S.

Note that this provides an alternative mechanism to a combination of nested JSON schema and JSON references [11].

## A.1.5 Schema Composition

A collection of schema can be structurally composed through arrays and objects to define new schema. These composition operators are required to define the schema of structured attributes, described in Section 2.6 above.

If $S_1$, ..., $S_n$ denote n schema then their *array composition* is the schema A = { `"type"`: `"array"`, `"items"`: [ $S_1$, ..., $S_n$] }. The new schema A describes array values of size k where each of its items is valid for the corresponding schema in the `"items"` field of A.

If $K_1$, ..., $K_n$ are strings representing keys then the *object composition* of the schema $S_1$, ..., $S_k$ with those strings is the schema O = { `"/K_1"`:$S_1$,..., `"/K_n"`:$S_n$ }. The new schema O describes object values that must contain the keys $K_1$, ..., $K_n$ and the values corresponding to each of these keys must be valid for $S_1$, ..., $S_n$ respectively.

For example, if S = { `"/age"`: `"$integer"` } and S' = `"$boolean"` are schema then their array composition is the schema { `"type"`: `"array"`; `"items"`: [ { `"/age"`: `"$integer"` }, `"$boolean"` ] } which validates values such as the array [ { `"age"`: 12 } , true ].

Similarly, the object composition of S and S' using the keys K = `"stats"` and K' = `"alive"` results in the following schema { `"/stats"`: { `"/age"`: `"$integer"` }, `"/alive"`: `"$boolean"` } which validates values such as the object { `"stats"`: { `"age"`: 321 }, `"alive"`: false }.

## A.2 Schema References and Resolution

The `$` in the above example denotes a *reference* and indicates that part of the schema being described is found elsewhere. A reference is always denoted by a string value beginning with `$`, and every string in a schema beginning with `$` is treated as a reference. The interpretation and resolution of a reference within a PSI schema is determined by its *address* and its *parameters*.

The string after the `$` is the reference's *address*. If a reference's address is a URI (i.e., begins with `http:` or some other URI scheme) then the address is said to be *global*, otherwise it is *local*. If a reference appears as a key in an object then it is a *parameterised reference* and the value associated with that key is the reference's parameters. Reference parameters are always bundled as a JSON object value. References that appear elsewhere in a schema are non-parameterised references.

Reference resolution is performed as part of the compilation process described in the next section. The compilation process manages a *resolution context* which is a map between local addresses and global addresses or schema. This context is always pre-populated with the names given in Appendix B. Once a local address has been resolved to a schema its resolution is complete. If a local address is resolved to a global address it is subsequently resolved like a global address. Global addresses are always resolved via GET requests to the address's URI.

Reference parameters are converted to a HTTP query as follows. Each property of the form `"key"`: `value` in the object holding the reference parameters is converted into a query fragment `key=evalue` where `evalue` is a URL encoded version of `value`. The query fragments are joined with `&` characters to form the final query string used in the GET request.

PSI schema references are an alternative mechanism to JSON references [11] and have the following key differences. First, referenced schema can be parameterised with a JSON object of name-value pairs, while a JSON reference is defined solely by the reference URI. Second, PSI references distinguish between a *local* namespace that exists during compilation, which contains predefined schema and locally-defined schema, and the *global* namespace of all full URIs.

## A.3 Compilation to JSON Schema

The PSI schema language can be "compiled" to JSON schema in the sense that each of the above constraints can be converted into one or more JSON schema properties that express the same constraints.

The following procedure describes how a PSI schema S can be translated into its corresponding JSON schema S', which will be valid against the JSON Hyper Schema draft version 4 [8]. Note that compilation of PSI schema for rich values results in a `mediaType` property being added to the corresponding JSON schema, which is outside the current JSON schema specification but which will be ignored by compliant validators. The resolution context C is assumed to be initialised with all of the predefined schema given in Appendix B. It is recommended that the key-value pair `"$schema"` : `"http://json-schema.org/draft-04/hyper-schema#"` be added to S' to assist JSON Schema validators in applying the correct validation rules.

Compile(S,C):

- If S is an integer, number, boolean then return S.
- If S is a string of the form "$R" then
    - Resolve the reference R using context C to get R' and return Compile(R',C)
    - Otherwise return S
- If S is a string of the form "@T" then return the JSON object { `"type"` : `"string"`, `"format"` : `"uri"`, `"mediaType"` : T }
- If S is an array then
    - Initialise an empty array S' and for each item I in S add the value Compile(I,C) to S'
    - Return S'
- If S is an object then initialise an empty JSON object S'
    - For each property P in S with key K of the form "#F" and value V
        - Add the pair (F,V) to the resolution context C
    - If S contains a key of the form "$R" which does not equal "$ref" or "$schema", with the object value V, then resolve the reference R in context C with arguments V to obtain R' and return Compile(R',C)
    - For each other property P in S with key K and value V
        - If K is `allItems` then add the property with key `items` and value Compile(V,C) to S'
        - If K is equal to "/*" then add the property with key `additionalProperties` and value Compile(V,C) to S'. Add the property `"type"`: `"object"` to S'
        - If K is of the form "/F", "/F=", "?F", "?F=" then
            - Add the property with key F to the object T associated with the key `"properties"` in S', creating a new property in S' with key `"properties"` and empty object value T if necessary.
            - If the key K ended with = then associate the value I = { `"enum"`: V } with F in T, otherwise associate the value I = Compile(V,C) with F in T.
            - If the key K started with / then add F to the array Q associated with the key `"required"` in S', creating a new property in S' with key `"required"` and empty array value Q if necessary.
            - Add the property `"type"`: `"object"` to S'.
        - Otherwise add the property K with value Compile(V,C) to S'
    - Return S'

## A.4 Schema Validation

PSI schema validation is performed by first compiling the schema into JSON schema and then using JSON schema validation semantics and implementations to carry out the validation. See the JSON Schema specification for details [6,7,8].

### A.4.1 Rich Value Validation

A (rich) value V may be validated against the PSI schema @T using the following algorithm:

- Compile PSI schema @T to JSON schema S and perform normal validation
- If V is valid against S (i.e., V is a URI) then
    - If the URI scheme is "http" then
        - V is valid if and only Content-Type is present and matches T.
    - Else if the URI scheme is "data" then
        - Extract the media type component of the data URI.
        - V is valid if and only if the media type matches T.
- V is not valid

# Appendix B: Predefined Schema

This appendix contains definitions of predefined schema that a PSI service must provide. Some schema are templates that accept, via GET, values for those parts that are variable. If a schema property's value is not provided, that property is absent from the schema object returned in a GET request. As described in Section 2.3 on schema resources, template arguments are preceded by the percent symbol (%). Values passed in via GET that are neither template arguments nor existing keys in the schema template are added as new properties.

All predefined schema resolve to a global address with relative URI `{schemaRoot}/{name}` where `{schemaRoot}` is the URI for the schema collection given in the PSI service's top-level description (see Section 2.2) and `{name}` denotes the name of the predefined schema. For example, if the service's schema collection is located at `http://example.org/schema`, then `$integer` resolves to `http://example.org/schema/integer`.

Each predefined schema is listed below with a brief description of its purpose and the template that is returned via a GET request. The examples below assume that the schema collection is located at `/schema`, although a particular PSI service may choose to locate this collection elsewhere (the service's top-level representation will provide the schema collection's URI).

## B.1 Value Schema

**`$integer`**

The integer schema only validates integer values. Its template can take arguments `min` to specify an inclusive minimum value, `max` to specify an inclusive maximum value, and `default` to specify a default value.

```
GET /schema/integer?template=true
----
200 OK
{
    "type": "integer",
    "minimum": "%min",
    "maximum": "%max",
    "default": "%default"
}
```

**`$number`**

The number schema only validates number values. Its template can take arguments `min` to specify an inclusive minimum value, `max` to specify an inclusive maximum value, and `default` to specify a default value.

```
GET /schema/number?template=true
----
200 OK
{
    "type": "number",
    "minimum": "%min",
    "maximum": "%max",
    "default": "%default"
}
```

**$boolean**

The boolean schema only validates boolean values. Its template can take the argument `default` to specify a default value.

```
GET /schema/boolean?template=true
----
200 OK
{
    "type": "boolean",
    "default": "%default"
}
```

**$string**

The string schema only validates string values. Its template can take the argument `default` to specify a default value.

```
GET /schema/string?template=true
----
200 OK
{
    "type": "string",
    "default": "%default"
}
```

**$object**

The object schema only validates object values. Its template can take the argument `default` to specify a default value.

```
GET /schema/object?template=true
----
200 OK
{
    "type": "object",
    "default": "%default"
}
```

**$array**

The array schema only validates array values. Its template can take the arguments: `items` to specify a schema that all items in the array must match, or an array of schema in which case matching values are described as in [Appendix A](). The `size` argument specifies the number of items a matching array value must contain.

```
GET /schema/array?template=true
----
200 OK
{
    "type":         "array",
    "items":        "%items",
    "minItems":     "%size",
    "maxItems":     "%size"
}
```

**`$atomicValue`**

> This schema only validates atomic values, that is, values that are integer, number, boolean, or string.
>
> ```
> GET http://example.org/schema/atomicValue
> ----
> 200 OK
> {
>     "type": [ "integer", "number", "boolean", "string" ]
> }
> ```

**`$atomicValueSchema`**

> This schema only validates *schema* that validate atomic values.
>
> ```
> GET http://example.org/schema/atomicValueSchema
> ----
> 200 OK
> {
>     "/type": { "enum" : [ "integer", "number", "boolean", "string" ] }
> }
> ```

**`$numberSchema`**

> This schema only validates *schema* that validate number values.
>
> ```
> GET http://example.org/schema/numberSchema
> ----
> 200 OK
> {
>     "/type": { "enum" : [ "integer", "number" ] }
> }
> ```

**`$nominalValueSchema`**

> This schema only validates *schema* that validate nominal values.
>
> ```
> GET http://example.org/schema/nominalValueSchema
> ----
> 200 OK
> {
>     "/enum":   { "$array": { "allItems": "$string" } }
> }
> ```

**`$uri`**

> This schema validates strings that are valid URIs according to RFC3986 [9].
>
> ```
> GET http://example.org/schema/uri
> ----
> 200 OK
> {
>     "type":    "string",
>     "format":  "uri"
> }
> ```

`$richValueSchema`

This schema only validates *schema* that validate rich value URIs for a specific media type.

```
GET http://example.org/schema/richValueSchema?template=true
----
200 OK
{
    "/type=":       "string",
    "/format=":     "uri",
    "/mediaType=":  "%mediaType"
}
```

## B.2 Resource Schema

`$relation`

This schema defines the structure of relation description response.

```
GET /schema/relation
----
200 OK
{
    "/psiType=":        "relation",
    "/uri":             "$uri",
    "?description":     "$string",
    "/size":            "$integer",
    "/defaultAttribute": "$uri",
    "/attributes":      { "$array" : { "items" : "$uri" } },
    "?querySchema":     "$object"
}
```

`$attribute`

A schema that is only valid for attribute description responses.

```
GET /schema/attribute
----
200 OK
{
    "/psiType=":        "attribute",
    "/uri":             "$uri",
    "?description":     "$string",
    "/emits":           "$object",
    "?relation":        "$uri",
    "?subattributes":   {
        "oneOf" : [
            { "$array": { "allItems": "$uri" } },
            { "/*" : "$uri" }
        ]
    },
    "?querySchema":     "$object"
}
```

**`$arrayAttribute`**

> A schema that only validates attributes with schema that define array values with items that match the schema passed in through the `allItems` argument.
>
> ```
> GET /schema/arrayAttribute?template=true
> ----
> 200 OK
> {
>     "allOf":   [ "$attribute" ],
>     "/emits": {
>         "/type=": "array",
>         "/items": { "$array": { "allItems": "%allItems" } }
>     }
> }
> ```

**`$numberAttribute`**

> A schema that only validates attributes with schema that define number values.
>
> ```
> GET /schema/numberAttribute?template=true
> ----
> {
>     "allOf" : [ "$attribute" ],
>     "/emits": {
>         "/type":   { "enum": [ "integer", "number" ] }
>     }
> }
> ```

**`$fixedAttribute`**

> A schema that only validates attributes with schema that only validate values from a fixed set of alternatives. This set of alternatives is provided via the `values` argument.
>
> ```
> GET /schema/fixedAttribute?template=true
> ----
> 200 OK
> {
>     "allOf":   [ "$attribute" ],
>     "/emits": {
>         "/enum=": "%values"
>     }
> }
> ```

**`$nominalAttribute`**

> A schema that only validates attributes with schema that define
> values that come from a fixed set of alternatives which are not specified in advance. The `allItems` argument can be used to specify what type of nominal values the attribute can return.
>
> ```
> GET /schema/nominalAttribute?template=true
> ----
> 200 OK
> {
>     "allOf":   [ "$attribute" ],
>     "/emits": {
>         "/enum": { "$array": { "allItems": "%allItems" } }
>     }
> }
> ```

**`$atomicAttribute`**

> A schema that only validates attributes with schema that only validate for atomic values.
>
> ```
> GET /schema/atomicAttribute?template=true
> ----
> 200 OK
> {
>     "allOf":    [ "$attribute" ],
>     "/emits": {
>         /type": { "enum": [ "integer", "number", "boolean", "string" ] }
>     }
> }
> ```

**`$richValueAttribute`**

> A schema that only validates attributes with schema that only validate rich values with a particular media type.
>
> ```
> GET /schema/richValueAttribute?template=true
> ----
> 200 OK
> {
>     "allOf":    [ "$attribute" ],
>     "/emits": { "$richValueSchema" : { "mediaType": "%mediaType" } }
> }
> ```

---

# Acknowledgements

We would like to thank Christfried Webers for his valuable feedback on earlier versions of this specification document.